\documentclass[aps,preprint,showpacs,amsmath,amssymb,prd,nofootinbib]{revtex4}
\usepackage{amssymb,graphics,graphicx}

\newcommand{\be}{\begin{equation}} \newcommand{\ee}{\end{equation}}

\begin{document}
\title
{\Large \bf LHC Higgs Production and Decay in the $T'$ Model }

\author{Paul H. Frampton$^{1}$\footnote{frampton@physics.unc.edu}, Chiu Man Ho$^{2}$\footnote{chiuman.ho@vanderbilt.edu}, Thomas W. Kephart$^{2}$\footnote{tom.kephart@gmail.com}, and Shinya Matsuzaki$^{3}$\footnote{synya@pusan.ac.kr}}
\affiliation{$^{1}$Department of Physics and Astronomy, University of North Carolina,
Chapel Hill, NC 27599-3255.\\$^{2}$Department of Physics and Astronomy, Vanderbilt University, Nashville, Tennessee 37235, USA.\\$^{3}$Department of Physics, Pusan National University, Busan 609-735, Korea. \\}

\date{\today}

\begin{abstract}
 At $\sqrt{s} = 7$ TeV, the standard model needs at least $10\, (fb)^{-1}$ integrated luminosity
 at LHC to make a definitive discovery of the Higgs boson.
Using binary tetrahedral ($T^{'}$) discrete flavor symmetry, we discuss how
 the decay of the lightest $T'$ Higgs into $\gamma\,\gamma$ can be effectively enhanced and
 dominate over its decay into $b \, \bar{b}$. Since the two-photon final state allows for a clean reconstruction,
 a decisive Higgs discovery may be possible at $7$ TeV with the integrated luminosity only
 of $\sim 1\, (fb)^{-1}$.
 \end{abstract}

\maketitle

\section{Introduction}

The standard model (SM) of particle physics is a well-tested theory which successfully predicts
the strong and electroweak interactions of elementary particles. While its predictive
power is impressive, it has limitations. First, in its minimal form, neutrinos have been introduced
as massless particles. However, a wealth of experimental data have confirmed that neutrinos are massive and that
flavors can mix. Undoubtedly, neutrino mixing is the first indisputable physics
beyond the minimal standard model. Moreover, in order to specify the standard model and make predictions, we need approximately 28 free parameters,
including the gauge couplings, quark and lepton masses, mixing angles and possible CP-violating phases, etc.

Grand unification theories (GUT), with or without supersymmetry (SUSY), have been invoked to explain the origin
of these free parameters or the relationship between them \cite{su5,so10,su5susy}. These models usually focus on reducing
the number of parameters in the gauge sector and either the quark or lepton sector, but not both.

A notable  alternative to GUTs
are models constructed with  discrete flavor symmetry. Here we will focus on the binary tetrahedral group $T'$, which provides calculability to both quark
and lepton sectors \cite{Frampton:1994rk,Frampton:2007et,Frampton:2008bz}. This model relates quarks and leptons through the
$T'$ symmetry whose irreducible representations are three singlets, three doublets and a triplet.
Since different quark
families are assigned to $T'$ singlets and doublets, mass hierarchies in the quark sector   appear  naturally in the quark sector.
Also, the fact that all the $SU_L(2)$ lepton doublets are assigned to a $T'$ triplet is to some extent a unification of the lepton sector.
The renormalizable $T'$ model has led to successful predictions of the tribimaximal neutrino mixing matrix as well as the Cabibbo angle \cite{Frampton:2007et,Frampton:2008bz}. Recently, it has been shown that the discrepancy between the SM prediction and experimental
value of muon $g-2$ factor can be easily accommodated in this model \cite{Ho:2010yp}. More details about the $T'$ model, its variants and
other related models can be found in the literature \cite{FamilySymmetryRefs}. The success of the renormalizable $T'$ model inspires us to ask
if it can be tested at the LHC. In this article, we
study  Higgs production and decay in the $T'$ model at the LHC.

Standard model Higgs production and decays have been studied in considerable detail~\cite{Spira:1997dg}. For instance, due to
the high gluon luminosity,
gluon-gluon fusion $g\,g \rightarrow h$ is the dominant Higgs production
mechanism at the LHC for Higgs masses up to $M_h\sim 1$ TeV. This is about an order of magnitude larger than the next most
important production process $ q \,\bar{q}' \rightarrow h \,W^{\pm} $. The gluon coupling to Higgs is mediated by the triangular quark loops,
and the process is dominated by the top and bottom loops. For $M_h \lesssim 160 $ GeV, the branching ratio for the decay process $h\rightarrow b \, \bar{b}$ dominates over all other decay processes such as $h\rightarrow g \, g$,\, $h\rightarrow W \, W^{\ast}$, \,$h\rightarrow Z \, Z^{\ast}$ and
$h\rightarrow \gamma\, \gamma $.  In this Higgs mass regime, the production of all other fermion pairs are relatively suppressed compared to $b\,\bar{b}$ because they are either produced by a relatively lower branching ratio through Higgs decay or by mixing.
For $M_h >160 \,{\rm GeV}$, the branching ratio for the decay process $h\rightarrow W W$ will take over and dominate.

Given the current limited luminosity of LHC at $\sqrt{s} = 7$ TeV, the more relevant mass range for the SM Higgs would be $M_h \lesssim 160 $ GeV. Due to the large QCD background, it is difficult to confirm the processes $h\rightarrow b \, \bar{b}$ and $h\rightarrow g \, g$. While $h\rightarrow W \, W^{\ast}$ and $h\rightarrow Z \, Z^{\ast}$ have relatively higher rates, the analysis is complicated by escaping neutrinos. As a consequence, in the regime $M_h \lesssim 160 $ GeV, the cleanest signal would be $h\rightarrow \gamma\, \gamma $ despite its tiny rate.

In this article, we focus on the decays of the lightest $T'$ Higgs with mass less than 160 GeV. We compare the decay rates of
$T'$ Higgses with those of the SM. In particular, we will show that in the fermiophobic limit, the decay of the lightest $T'$ Higgs into $\gamma\,\gamma$ is effectively enhanced and dominates over its decay into $b \, \bar{b}$. Since the high $p_\textrm{T}$ two-photon final state allows a clean reconstruction, a decisive Higgs discovery may be possible in this limit, even at $\sqrt{s} = 7$ TeV with the integrated luminosity of only  $\sim 1(fb)^{-1}$. As a bonus, the lightest $T'$ Higgs can be unambiguously distinguished from the SM Higgs.

\section{The $T^{'}$ Model and lightest Higgs boson }

We start with a brief review of the simplified model proposed in~\cite{Frampton:2008bz}
based on the global symmetry $(T^{'} \times Z_2)$. In particular, we will ignore the lepton sector which
will not be relevant to our study in this article.
Interested readers can refer to
\cite{Frampton:1994rk,Frampton:2007et,Frampton:2008bz} for more details.

In the $T'$ model, left-handed quark doublets \noindent $(t, b)_L, (c, d)_L, (u, d)_L$
are assigned under this global symmetry as
\begin{equation}
\begin{array}{cc}
\left( \begin{array}{c} t \\ b \end{array} \right)_{L}
~ {\cal Q}_L ~~~~~~~~~~~ ({\bf 1_1}, +1)   \\
\left. \begin{array}{c} \left( \begin{array}{c} c \\ s \end{array} \right)_{L}
\\
\left( \begin{array}{c} u \\ d \end{array} \right)_{L}  \end{array} \right\}
Q_L ~~~~~~~~ ({\bf 2_1}, +1),
\end{array}
\label{qL}
\end{equation}
and the six right-handed quarks as
\begin{equation}
\begin{array}{c}
t_{R} ~~~~~~~~~~~~~~ ({\bf 1_1}, +1)   \\
b_{R} ~~~~~~~~~~~~~~ ({\bf 1_2}, -1)  \\
\left. \begin{array}{c} c_{R} \\ u_{R} \end{array} \right\}
{\cal C}_R ~~~~~~~~ ({\bf 2_3}, -1)\\
\left. \begin{array}{c} s_{R} \\ d_{R} \end{array} \right\}
{\cal S}_R ~~~~~~~~ ({\bf 2_2}, +1).
\end{array}
\label{qR}
\end{equation}
 The quark-Yukawa sector in the model is thus given as~\cite{Frampton:2008bz}
\begin{eqnarray}
{\cal L}_Y^q
&=& Y_t ( \{{\cal Q}_L\}_{\bf 1_1}  \{t_R\}_{\bf 1_1} H_{\bf 1_1})
+ Y_b (\{{\cal Q}_L\}_{\bf 1_1} \{b_R\}_{\bf 1_2} H_{\bf 1_3} ) \nonumber \\
&&
+ Y_{{\cal C}} ( \{ Q_L \}_{\bf 2_1} \{ {\cal C}_R \}_{\bf 2_3} H^{'}_{\bf 3})
+ Y_{{\cal S}} ( \{ Q_L \}_{\bf 2_1} \{ {\cal S}_R \}_{\bf 2_2} H_{\bf 3})
\nonumber \\
&&
+ {\rm h.c.}.
\end{eqnarray}

\subsection{$T'$-Higgs couplings}

We focus on flavor-diagonal interactions and fermion couplings to a set of
neutral $T'$-Higgs bosons, $\{ H_r^{(i)} \}$, where $r=(1_1, 1_3, 3, 3')$ denotes
the $T'$-irreducible representation and $i$ denotes components in the $T'$-multiplet\footnote{As was studied in the model building of Ref.[8], if we go beyond the mininal $T'$ model to incorporate mixing with the third generation of quarks, we may encounter flavor changing neutral current (FCNC) problems because the Higgs bosons can have off-diagonal flavor couplings, unlike the standard model Higgs. Then, the size of $T'$ Yukawa couplings to the mass-eigenstate Higgs bosons, namely $Y_f^2 (a_n^f)^2 /M_{H_n}^2  \propto  (a_n^f/R_f)^2 (g_{hff}^{\rm SM})^2 M_{H_n}^2$, would be constrained by the FCNC issue, which may give further constraints on the parameters $a_n^f$ and $R_f$. More on this issue is beyond scope of the present article and is to be pursued in detail in the future.}. Let the Higgs vacuum expectation values (VEVs) be $\langle H_r^{(i)} \rangle = v_r^{(i)}/\sqrt{2}$.
 Expanding fields in terms of the Clebsch-Gordan coefficients~\cite{Frampton:2008bz},
 we have
\begin{eqnarray}
{\cal L}_Y^q \Bigg|_{\rm flavor-diagonal}^{\rm neutral}
 &=&
Y_t \bar{t} t \frac{(H_{1_1} + v_{1_1})}{\sqrt{2}} +
Y_b \bar{b} b \frac{(H_{1_3} + v_{1_3})}{\sqrt{2}}
\nonumber \\
&&
- \frac{Y_{\cal C}}{\sqrt{6}} \bar{c} c  \frac{(H_{3'}^{(1)} + v_{3'}^{(1)})}{\sqrt{2}}
+ \frac{Y_{\cal S}}{\sqrt{3}} \bar{s} s \frac{(H_3^{(1)} + v_3^{(1)})}{\sqrt{2}}
\nonumber \\
&&
+ \frac{Y_{\cal S}}{\sqrt{6}}\bar{d} d \frac{(H_{3}^{(1)} + v_{3}^{(1)})}{\sqrt{2}}
+ \sqrt{\frac{2}{3}} Y_{\cal C}  \bar{u} u \frac{(H_{3'}^{(2)} + v_{3'}^{(2)})}{\sqrt{2}}
\,.
\end{eqnarray}
The relevant Yukawa couplings and fermion masses are thus read off:
\begin{eqnarray}
g_{H_{1_1} tt} &=& Y_t  \,, \qquad m_t = \frac{Y_t}{\sqrt{2}} v_{1_1}
\,, \\
g_{H_{1_3} bb} &=& Y_b \,, \qquad m_b = \frac{Y_b}{\sqrt{2}} v_{1_3}
\,, \\
g_{H_{3'}^{(1)} cc} &=& \frac{Y_{\cal C}}{\sqrt{6}} \,, \qquad m_c = \frac{Y_{\cal C}}{2 \sqrt{3}} v_{3'}^{(1)}
\,, \\
g_{H_3^{(1)} ss} &=& \frac{Y_{\cal S}}{\sqrt{3}} \,, \qquad m_s =\frac{Y_{\cal S}}{\sqrt{6}} v_3^{(1)}
\,, \\
g_{H_{3}^{(1)} dd} &=& \frac{Y_{\cal S}}{\sqrt{6}} \,, \qquad m_d =  \frac{Y_{\cal S}}{2 \sqrt{3}} v_{3}^{(1)}
\,, \\
g_{H_{3'}^{(2)} uu} &=& \sqrt{\frac{2}{3}} Y_{\cal C} \,, \qquad m_u = \frac{Y_{\cal C}}{\sqrt{3}} v_{3'}^{(2)}
\,.
\end{eqnarray}
Fixing fermion masses to be the same as those in the standard model, namely
$m_f = \frac{g_{hff}^{\rm SM}}{\sqrt{2}} v_{\rm EW}$,
we may express the $T'$-Yukawa couplings comparing those in the standard model
to get
\begin{eqnarray}
 g_{H_{1_1} tt} &=& \left( \frac{v_{\rm EW}}{v_{1_1}} \right) g_{h tt}^{\rm SM}
 \,, \label{gHtt} \\
  g_{H_{1_3} bb} &=& \left( \frac{v_{\rm EW}}{v_{1_3}} \right) g_{h bb}^{\rm SM}
 \,, \\
  g_{H_{3'}^{(1)} cc} &=& \left( \frac{v_{\rm EW}}{v_{3'}^{(1)}} \right) g_{h cc}^{\rm SM}
 \,, \\
  g_{H_{3}^{(1)} ss} &=& \left( \frac{v_{\rm EW}}{v_{3}^{(1)}} \right) g_{h ss}^{\rm SM}
\,, \\
  g_{H_{3}^{(1)} dd} &=& \left( \frac{v_{\rm EW}}{v_{3}^{(1)}} \right) g_{h dd}^{\rm SM}
\,, \\
  g_{H_{3'}^{(2)} uu} &=& \left( \frac{v_{\rm EW}}{v_{3'}^{(2)}} \right) g_{h uu}^{\rm SM}
\,. \label{gHuu}
\end{eqnarray}

We next turn to the gauge-Higgs sector,
${\cal L}_{\rm GH} = \sum_{r,i} |D_\mu H_r^{(i)}|^2$,
where
all the $T'$-Higgs fields couple to the electroweak gauge bosons.
 The $W$ and $Z$ boson masses are thus expressed in terms of
the $T'$-Higgs VEVs as follows:
\begin{equation}
  M_W^2 = \frac{g_W^2}{4} \sum_{r,i} (v_r^{(i)})^2
\,, \qquad
M_Z^2 = \frac{M_W^2}{c_W^2}
\,,
\end{equation}
where $g_W$ is the $SU(2)_W$ gauge coupling and
$c_W = \frac{g_W}{\sqrt{g_W^2 + g_Y^2}}$ with $g_Y$ being $U(1)_Y$ gauge coupling~\footnote{
Note that we have $\rho = 1$ at tree level as in the standard model.
This is because $T'$-symmetry commutes with the electroweak symmetry as well as
the custodial symmetry. }.
  We fix the $W$ and $Z$ boson masses
to be those in the standard model. This can be achieved by identifying
the electroweak scale $v_{\rm EW}$ as
\begin{equation}
 v_{\rm EW}^2 =  \sum_{r,i} (v_r^{(i)})^2
\,,
\end{equation}
so that we have $M_W^2 = g_W^2 v_{\rm EW}^2/4 $.

The $T'$-Higgs couplings to $WW$ and $ZZ$ read
\begin{equation}
 {\cal L}_{H_r^{(i)} VV}
 = g_{H_r^{(i)} WW} H_r^{(i)} W^+_\mu W^{\mu -}
 + \frac{1}{2} g_{H_r^{(i)} ZZ} H_r^{(i)} Z_\mu Z^\mu
 \,,
\end{equation}
where
\begin{eqnarray}
 g_{H_r^{(i)} WW} &=&
g_W^2 v_r^{(i)} = \left( \frac{v_r^{(i)}}{v_{\rm EW}} \right) g_{hWW}^{\rm SM}
\,, \label{gHWW} \\
 g_{H_r^{(i)} ZZ} &=&
\frac{g_W^2}{c_W^2} v_r^{(i)} = \left( \frac{v_r^{(i)}}{v_{\rm EW}} \right) g_{hZZ}^{\rm SM}
\,, \label{gHZZ}
\end{eqnarray}
with $g_{hVV}^{\rm SM} $ ($V=W,Z$) being the corresponding coupling to the Higgs boson
in the standard model,
\begin{equation}
g_{hVV}^{\rm SM}= 4\,\frac{ M_V^2}{v_{\rm EW}}\,.
\end{equation}

\subsection{The lightest Higgs boson and its couplings}

Electroweak interactions mix $T'$-Higgs doublets.
Given an explicit form of Higgs potential, we can solve
such a mixing to get a set of mass-eigenstates $\{ H_n \}$
with their eigenfunctions, $a_n^r$.
Without knowing the explicit expression of Higgs potential,
in general, we may write
\begin{equation}
 H_r^{(i)} = \sum_n a_n^{(r,i)} H_n
 \,,
\end{equation}
where the expansion coefficients $a_n^{(r,i)}$ form an orthonormal complete set,
\begin{equation}
\sum_{r, i} a_n^{(r, i)} a_m^{(r, i)} = \delta_{nm}
\,, \label{ortho}
\end{equation}
which followed from the normalization
condition of the kinetic terms of $\{ H_{n} \}$.

Assuming a mass hierarchy for the Higgs bosons, $M_{H_0}<M_{H_1}< \cdots$, we can identify
the lightest Higgs boson as $H_0$ with mass, say, $\lesssim m_t\simeq 172$ GeV.
Hereafter we shall confine ourselves to the phenomenology of this $H_0$.

It is convenient to introduce a ratio,
\begin{equation}
 R_r^{(i)} = \frac{v_r^{(i)}}{v_{\rm EW}}
\,,
\end{equation}
which satisfies
\begin{equation}
\sum_{r, i} (R_r^{(i)})^2 = 1.
\label{constraint}
\end{equation}
From  Eqs.(\ref{gHtt})-(\ref{gHuu}) and (\ref{gHWW})-(\ref{gHZZ}),
we then obtain the $H_0$ couplings to fermions,
\begin{eqnarray}
   g_{H_0 tt} &=& \left( \frac{a_0^{1_1}}{R_{1_1}} \right) g_{h tt}^{\rm SM}
 \,, \label{gH0tt} \\
  g_{H_0 bb} &=& \left( \frac{a_0^{1_3}}{R_{1_3}} \right) g_{h bb}^{\rm SM}
 \,, \label{gH0bb}\\
  g_{H_0 cc} &=& \left( \frac{a_0^{(3', 1)}}{R_{3'}^{(1)}} \right) g_{h cc}^{\rm SM}
 \,, \\
  g_{H_0 ss} &=& \left( \frac{a_0^{(3, 1)}}{R_{3}^{(1)}} \right) g_{h ss}^{\rm SM}
\,, \\
  g_{H_0dd} &=& \left( \frac{a_0^{(3,1)}}{R_{3}^{(1)}} \right) g_{h dd}^{\rm SM}
\,, \\
  g_{H_0 uu} &=& \left( \frac{a_0^{(3', 2)}}{R_{3'}^{(2)}} \right) g_{h uu}^{\rm SM}
\,, \label{gH0uu}
\end{eqnarray}
and gauge bosons
\begin{eqnarray}
 g_{H_0 VV} = \sum_{r,i} \left( a_0^{(r, i)} R_r^{(i)} \right) g_{hVV}^{\rm SM}
\,, \label{gH0VV}
\end{eqnarray}
where $V=W,Z$.

\section{LHC Higgs Decay and Production}

In this section, we study the decay modes of a light Higgs boson with mass in the range
109 GeV $\lesssim M_{H_0} \lesssim (2 M_W \simeq) 160$ GeV~\footnote{
The lower bound comes from the exclusion limit of the direct search of
$h \to \gamma\gamma$ at the LEP II~\cite{LEPII}. }.
In this mass range, $h \to b \bar{b}$ and $h \to gg$ are dominant modes in the standard model,
where the top and bottom loops give the significant effect on the $h \to gg$ mode.
Here we shall focus on the top and bottom contributions to decay modes of the $T'$-Higgs boson $H_0$.
The relevant formulas for its partial decay widths of $H_0$ are given in Appendix~\ref{decay:formulas}.

From Eqs.(\ref{gH0tt})-(\ref{gH0uu}) and Eq.(\ref{gH0VV}), we see that
 the difference between the standard model and the $T'$ model is handled by
 two kinds of parameters, $a_0^{(r, i)}$ and $R_{r}^{(i)}$.
Since we are interested in the top and bottom contributions,
we may take $a_0^{(r,i)} =0$ except $a_0^{1_1}$ and $a_0^{1_3}$ in Eqs.(\ref{gH0tt})-(\ref{gH0uu})
and keep only $R_{1_1}$ and $R_{1_3}$ nonzero in Eq.(\ref{gH0VV}), so that
all the Yukawa couplings other than those of top and bottom vanish and
the $H_0$-$V$-$V$ coupling is saturated by only $H_{1_1}$ and $H_{1_3}$ in the sum.
Note that Eqs.(\ref{ortho}) and (\ref{constraint}) then constrain the remaining parameters:
\begin{eqnarray}
  (a^{1_1}_0)^2 + (a_0^{1_3})^2 &=& 1
  \,, \qquad
0 \le  a_0^{1_1} \le 1 \,,
\qquad
 0 \le  a_0^{1_3} \le 1 \,,
\nonumber \\
 (R_{1_1})^2 + (R_{1_3})^2 &=& 1
 \,,
 \qquad
 0 \le  R_{1_1} \le 1 \,,
 \qquad
0 \le  R_{1_3} \le 1 \,.
\label{constraint2}
\end{eqnarray}

From Eqs.(\ref{gH0tt}), (\ref{gH0bb}), (\ref{gH0VV}) and Eq.(\ref{constraint2}),
one can see that the standard model limit is given by
\begin{equation}
  a_0^{1_3} \to R_{1_3} \to 0
  \,,
\end{equation}
 in such a way that $g_{H_0tt} \to g_{htt}^{\rm SM}$,  $g_{H_0bb} \to g_{hbb}^{\rm SM}$ and $g_{H_0VV} \to g_{hVV}^{\rm SM}$.
On the other hand, a fermiophobic limit can be taken as
\begin{equation}
 a_0^{1_3} \to 0
 \,, \label{fl}
\end{equation}
in a sense that the bottom Yukawa coupling goes to zero and $H_0 \to b \bar{b}$ mode gets highly suppressed to be zero,
while the top Yukawa coupling remains nonzero.

\begin{figure}
\begin{center}
 \includegraphics[width=5.95cm]{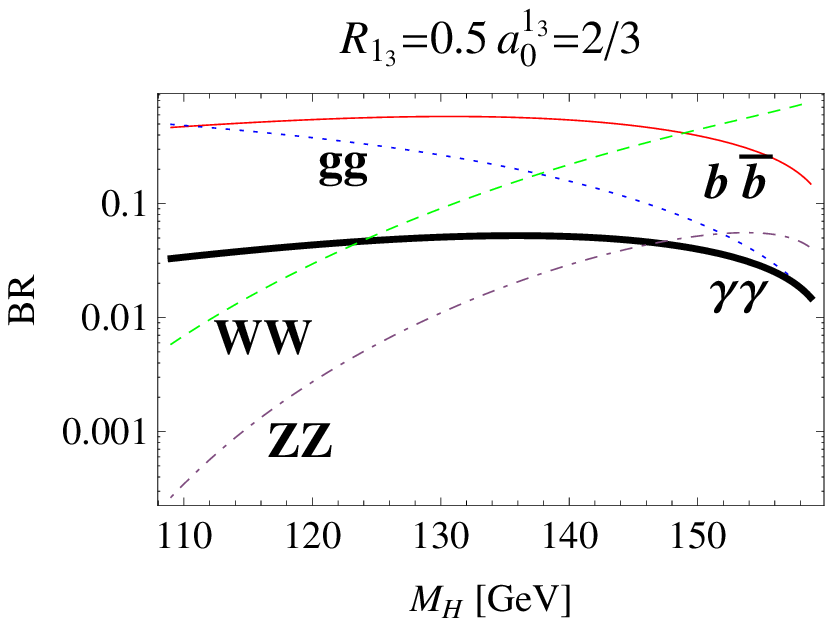}
 \includegraphics[width=5.95cm]{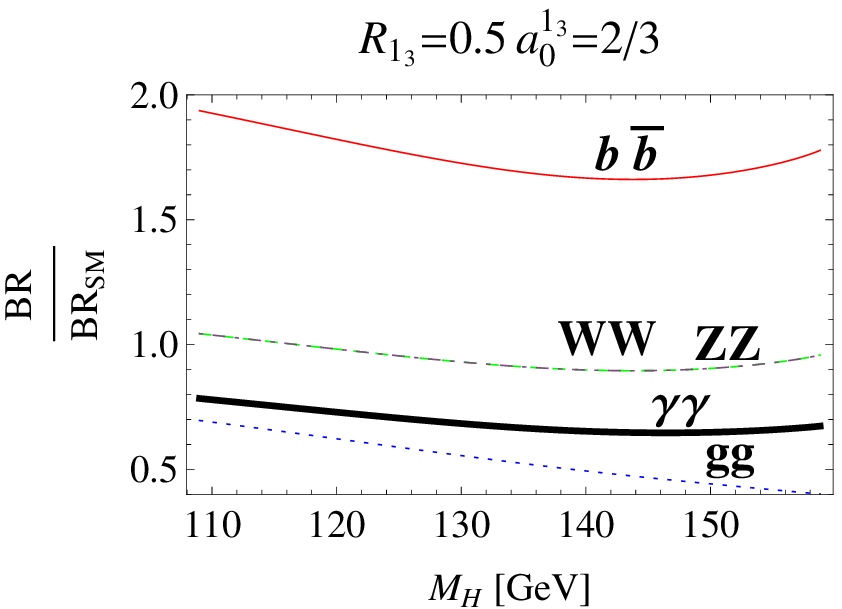}
 \caption{\footnotesize
Branching fraction of the lightest $T'$-Higgs boson
with $R_0^{1_3}=0.5$ and $a_0^{1_3}=2/3$.
}
\label{fig:br:1}
\end{center}
 \end{figure}

\begin{figure}
\begin{center}
 \includegraphics[width=5.95cm]{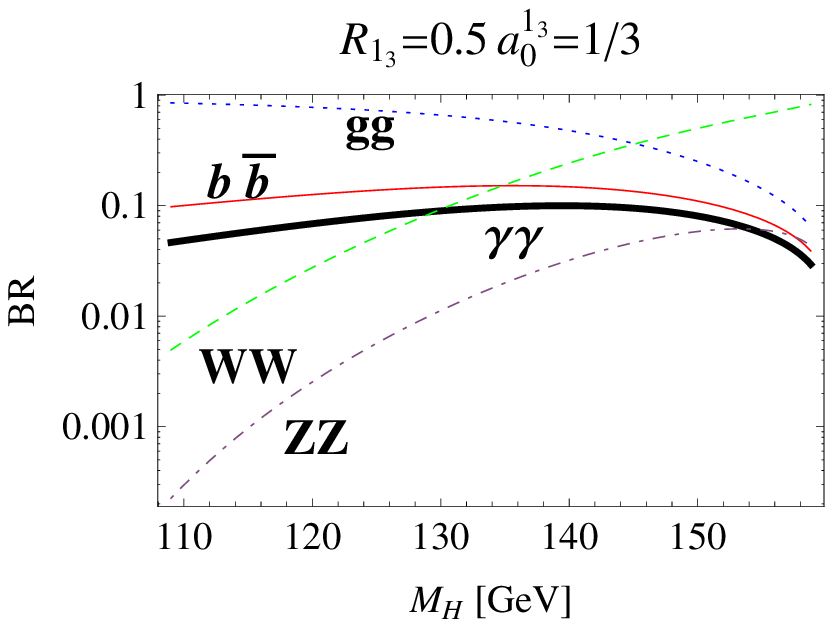}
 \includegraphics[width=5.95cm]{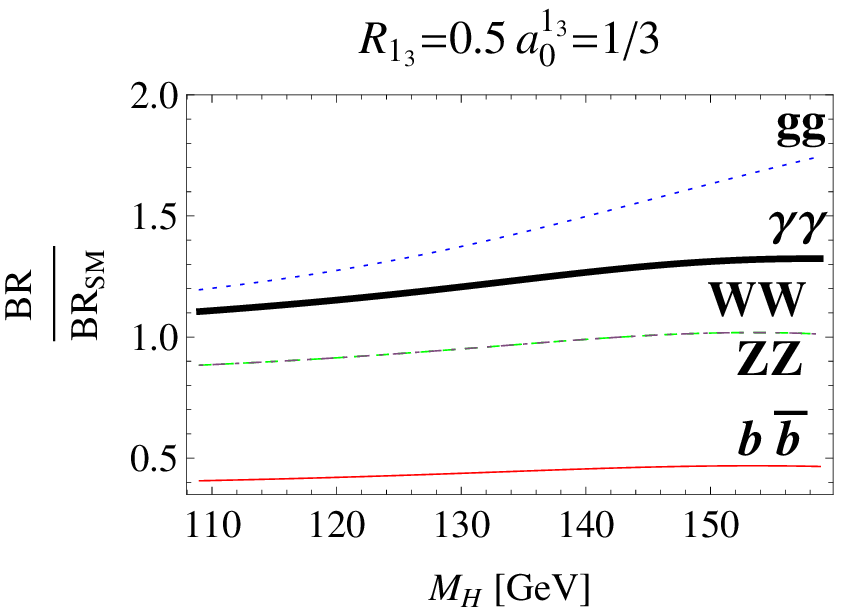}
 \caption{\footnotesize
Branching fraction of the lightest $T'$-Higgs boson
with $R_0^{1_3}=0.5$ and $a_0^{1_3}=1/3$. }
\label{fig:br:2}
\end{center}
 \end{figure}

\begin{figure}
\begin{center}
 \includegraphics[width=5.95cm]{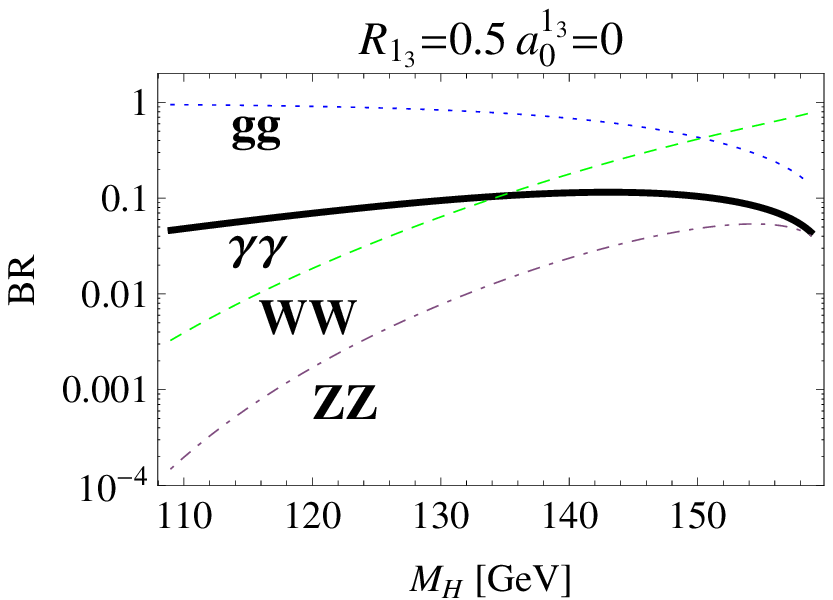}
 \includegraphics[width=5.95cm]{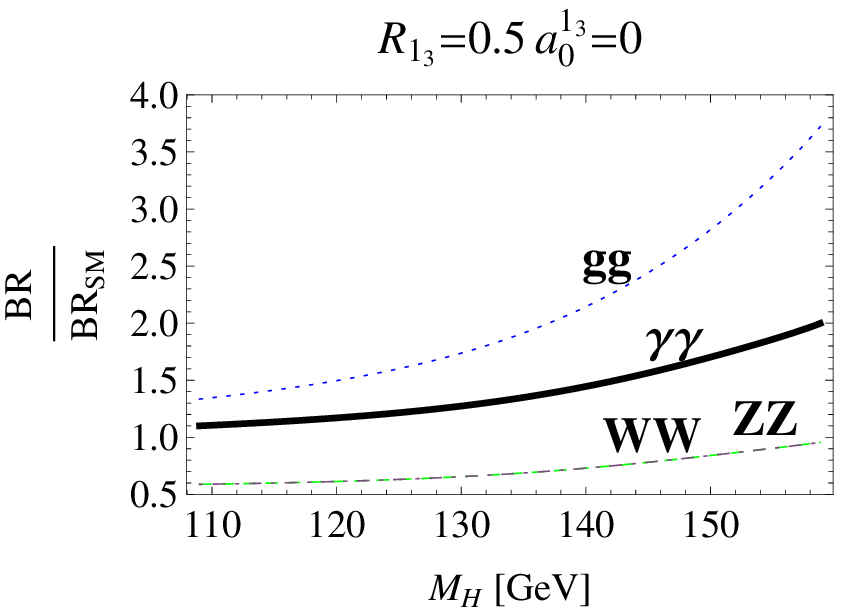}
 \caption{\footnotesize
Branching fraction of the lightest $T'$-Higgs boson
with $R_0^{1_3}=0.5$ and $a_0^{1_3}=0$ in which case the Higgs boson does not
couple to $b \bar{b}$ since the Yukawa coupling vanishes. }
\label{fig:br:3}
\end{center}
 \end{figure}

In Figs.~\ref{fig:br:1}-\ref{fig:br:3}, we show the
branching fraction of the lightest T' Higgs decay (left panels) and the ratio to that of the standard model Higgs
(right panels).
As a sample, we have taken $a^{1_3}_0=2/3, 1/3, 0$ with $R_{1_3}=0.5$  fixed
which monitors the interpolation between the standard model case and the fermiophobic case.
Figures~\ref{fig:br:1}-\ref{fig:br:3} imply that as $a_0^{1_3}$ approaches the fermiophobic limit $a_0^{1_3} \to 0$,
$H_0 \to \gamma\gamma$ becomes dominant 
in contrast to the case of the standard model Higgs in which $h \to b \bar{b}$ is dominant.
 Note that the fermiophobicity does not affect $WW$ and $ZZ$ decay modes so much (See Fig.~\ref{fig:br:3})
 since $g_{H_0 VV}/g_{hVV}^{\rm SM}=(\sqrt{1-(a_0^{1_3})^2} \sqrt{1-R_{1_3}^2} + a_0^{1_3} R_{1_3}) \to  \sqrt{1-R_{1_3}^2}$ when
 $a_0^{1_3} \to 0$.

\begin{figure}
\begin{center}
 \includegraphics[scale=0.8]{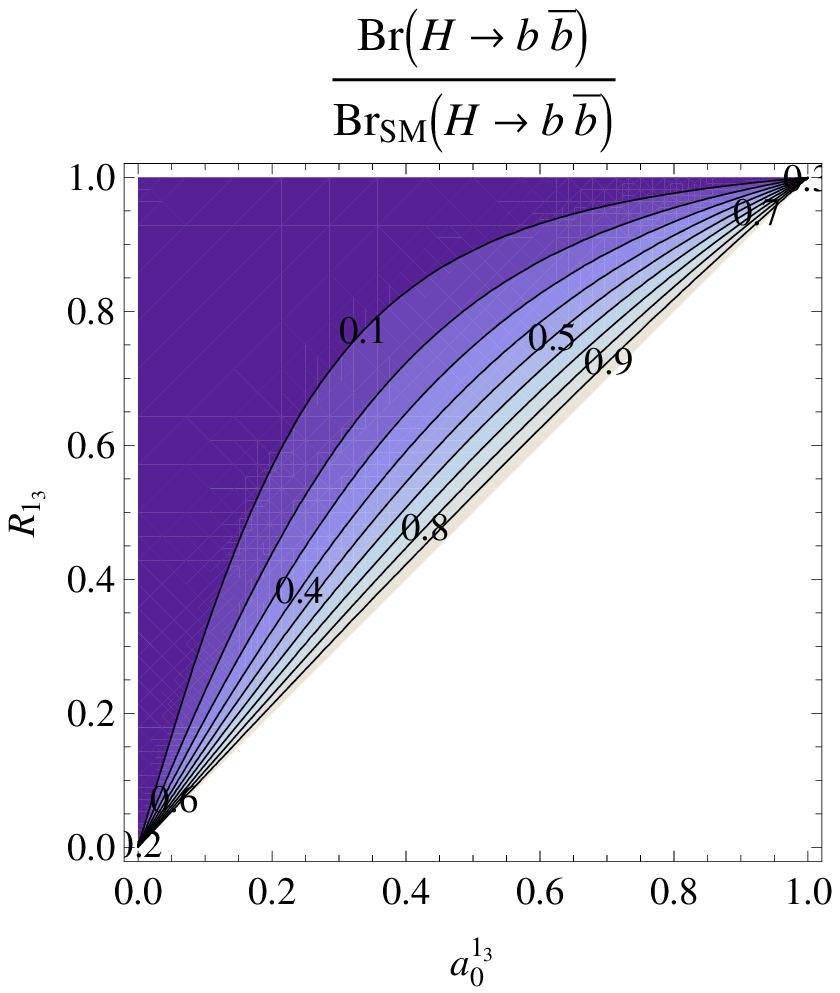}
 \caption{\footnotesize
Contour plot of ${\rm Br}(H_0 \to b \bar{b})/{\rm Br}_{\rm SM}(H_0 \to b \bar{b})$
on the ($a^{1_3}_0, R_{1_3}$) plane
for $M_H =120$ GeV. The contour points are restricted to a region where the value is less than
1. The limit $a_0^{1_3} \to R_{1_3} \to 0$ along the line $R_{1_3}=a_0^{1_3}$ corresponds to the standard model case.
}
\label{fig:cont:1}
\end{center}
 \end{figure}

\begin{figure}
\begin{center}
 \includegraphics[scale=0.8]{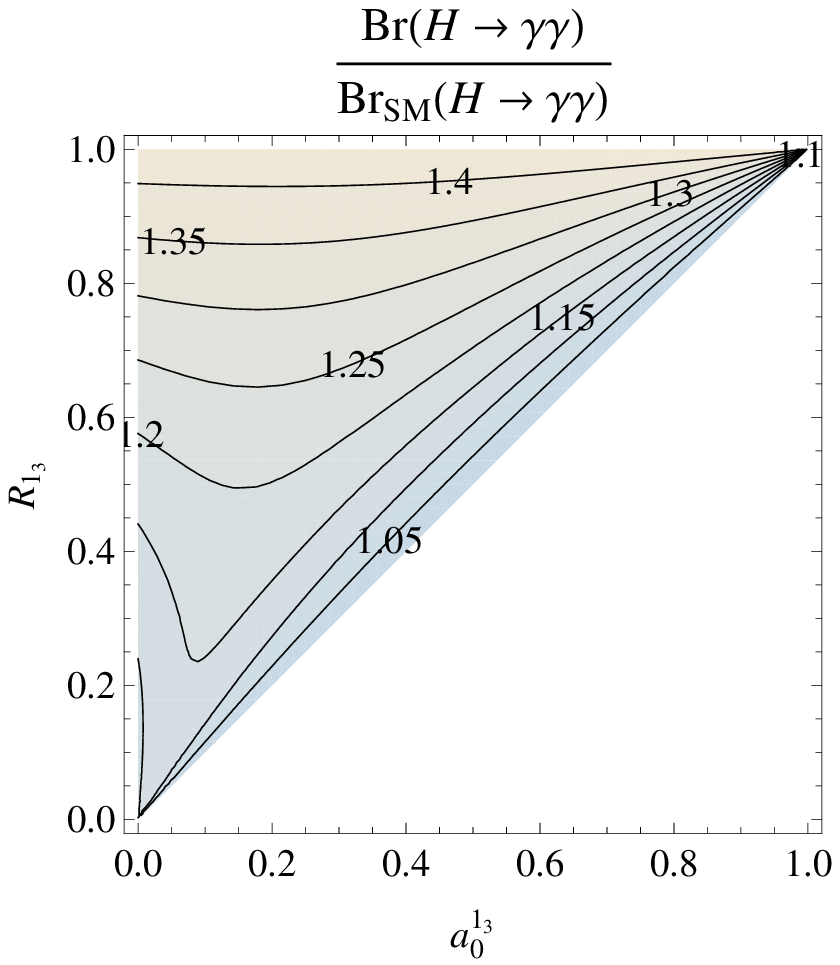}
 \caption{\footnotesize
Contour plot of ${\rm Br}(H_0 \to \gamma\gamma)/{\rm Br}_{\rm SM}(H_0 \to \gamma\gamma)$
on the ($a^{1_3}_0, R_{1_3}$) plane
for $M_H =120$ GeV. The contour points are restricted to a region where the value is larger than
1. The limit $a_0^{1_3} \to R_{1_3} \to 0$ along the line $R_{1_3}=a_0^{1_3}$ corresponds to the standard model case.
}
\label{fig:cont:2}
\end{center}
 \end{figure}

\begin{figure}
\begin{center}
 \includegraphics[scale=0.8]{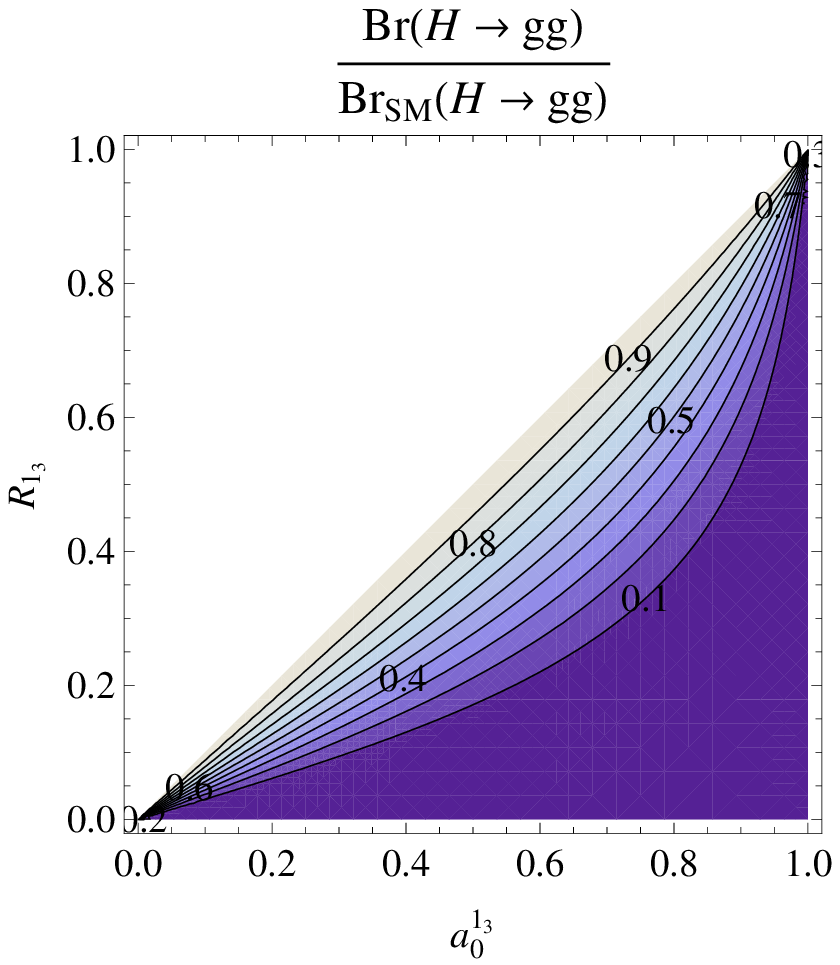}
 \caption{\footnotesize
Contour plot of ${\rm Br}(H_0 \to gg)/{\rm Br}_{\rm SM}(H_0 \to gg)$
on the ($a^{1_3}_0, R_{1_3}$) plane
for $M_H =120$ GeV. The contour points are restricted to a region where the value is less than
1. The limit $a_0^{1_3} \to R_{1_3} \to 0$ along the line $R_{1_3}=a_0^{1_3}$ corresponds to the standard model case.
} \label{fig:cont:3}
\end{center}
 \end{figure}

In Figs.~\ref{fig:cont:1}-\ref{fig:cont:3}, we show contour plots of ${\rm Br}(H_0 \to b \bar{b})$,
${\rm Br}(H_0 \to \gamma \gamma)$ and ${\rm Br}(H_0 \to gg)$ for $M_H = 120$ GeV in the entire region
of the parameter space $(a_0^{1_3}, R_{1_3})$ comparing with those of the standard model Higgs.
It is interesting to note from Figs.~\ref{fig:cont:1}-~\ref{fig:cont:3} that
$H_0 \to b \bar{b}$ mode is necessarily  suppressed when $H_0 \to \gamma \gamma$ mode
is enhanced, while $H_0 \to gg$ mode is enhanced at the same time which is due to
the remaining sizable top loop contribution: One cannot make both top and bottom quarks decoupled simultaneously
because of the constraint (\ref{constraint2}).

Finally, let us briefly discuss Higgs production.
In particular, in the true fermiophobic limit, which is realized by taking
$ a_0^{1_3} \rightarrow 0$, we find $ a_0^{1_1} \rightarrow 1$.
This implies that the gluon-gluon fusion through
the top-triangular loop will be the dominant production process for the fermiophobic $T'$-Higgs,
which is the same (within a percent) as  in the standard model. Even though the dominant Higgs production process is the same, the
fermiophobic $T'$-Higgs can be unambiguously distinguished from the SM Higgs at the LHC because $H_0 \rightarrow \gamma \, \gamma$
dominates over $H_0 \rightarrow b \, \bar{b}$ in the range $109$ GeV $\lesssim M_{H_0} \lesssim 160$ GeV. Since
$H_0 \rightarrow \gamma \, \gamma$ allows for a clean reconstruction, a decisive Higgs discovery may be
possible even at $\sqrt{s} = 7$ TeV with the integrated luminosity only of $\sim 1\,(fb)^{-1}$.

\section{Discussion}

\bigskip

\noindent
This article may be taken as a warning to experimentalists that the
properties of the lightest Higgs boson can readily depart very
significantly from the predictions of the minimal SM
with only one Higgs doublet, and with its 28 parameters
unconstrained by any further theoretical input.

\bigskip

\noindent
We have studied a model with a $(T^{'} \times Z_2)$ flavor
symmetry which commutes with the SM gauge group,
and which leads to agreement with the mixing matrices for neutrinos
and quarks. It necessarily changes the couplings, of the lightest
Higgs to the quarks and leptons, which are no longer
simply proportional to the fermion masses. This aspect of
the SM is its most fragile prediction.

\bigskip

\noindent
A similar, but different, illustration of this fragility
is provided by the variant axion model \cite{Chen:2010su}
motivated by, instead, solution of the strong CP problem.
In both cases, the delimiting of the SM parameters
changes the Yukawa sector.

\bigskip

\noindent
In the present case, the $T^{'}$ flavor symmetry can give rise to
optimism that the discovery of the Higgs may be expedited
because the product of the production cross-section and
the decay branching ratio is enhanced. As can be seen from
Figs. 3 and 5, the decay $H \rightarrow \gamma\gamma$
is generically larger even for $M_H = 120$ GeV
and becomes more so at larger Higgs mass. Even
with $\sqrt{s} = 7$ TeV, and 1 $(fb)^{-1}$, the LHC could
make a Higgs discovery.

\bigskip

\noindent
Many aspects of the SM have been confirmed to high
accuracy. These checks are principally
for the gauge sector which has a significant
geometrical underpinning and hence uniqueness.
The Yukawa sector, where most of the 28 free
parameters lie, does not have a geometrical
interpretation. The objective of the flavor symmetry
is to supply an explanation of
some of the parameters, and it is therefore
interesting to explore other predictions
for production and decay of Higgs at LHC.

\bigskip

\section*{Acknowledgments}

The work of P.H.F. was supported in part by
U.S. Department of Energy Grant No. DE-FG02-05ER41418.
C.M.H. and T.W.K.
were supported by US DOE grant DE-FG05-85ER40226.
S.M. was supported by the Korea Research
Foundation Grant funded by the Korean Government (KRF-2008-341-C00008).
P.H.F. and T.W.K. thank the Aspen Center for Physics for hospitality while this work was in progress.

\appendix
\renewcommand\theequation{\Alph{section}.\arabic{equation}}

\section{Formulas for Higgs decay widths}
 \label{decay:formulas}

In this appendix we shall present formulas of decay widths
relevant to the $H_0$-decay modes.

\subsection{$H_0 \to q \bar{q}$ mode}

In the standard model
the $h \to q \bar{q}$ decay width is calculated at the leading order of perturbation to be 
\begin{equation}
  \Gamma^{\rm SM}[ h \to q \bar{q}]
  = \frac{N_c M_h}{16 \pi} (g_{h qq}^{\rm SM})^2 \left( 1 - \frac{4 m_q^2}{M_h^2} \right)^{3/2}
  \,.
\end{equation}
To get the corresponding formula for $H_0 \to q \bar{q}$,
all we need to do is replace the Yukawa coupling and the Higgs boson mass with the appropriate ones.
Thus we have
\begin{equation}
  \Gamma^{T'}[ H_0 \to q \bar{q}]
  = \frac{N_c M_{H_0}}{16 \pi} (g_{H_0 qq})^2 \left( 1 - \frac{4 m_q^2}{M_{H_0}^2} \right)^{3/2}
  \,. \label{decay:Hqq}
\end{equation}

\subsection{$H_0 \to gg$ mode}

In the standard model
we compute the $h \to g g$ decay width at the leading order of perturbation to get
\begin{equation}
  \Gamma^{\rm SM} [h \to gg] =
  \frac{N_c^2 \alpha_S^2 M_h^3}{576 \pi^3}
  \Bigg|
     \sum_q \frac{g_{hqq}^{\rm SM}}{m_q} \left( 1 + (1-\tau_q) f(\tau_q)\right) \tau_q
  \Bigg|^2
  \,,
\end{equation}
where $\tau_q=4m_q^2/M_h^2$ and defined~\cite{Spira:1997dg}
\begin{eqnarray}
  f(\tau) = \Bigg\{
\begin{array}{cc}
  \left(\sin^{-1} \frac{1}{\sqrt{\tau}} \right)^2 & \tau \ge 1 \nonumber \\
   - \frac{1}{4} \left( \log \left( \frac{1+ \sqrt{1- \tau}}{1- \sqrt{1-\tau}} - i \pi \right)  \right)^2 & \tau <1
\end{array}
\,.
\end{eqnarray}
Replacing $g_{h qq}^{\rm SM}$ with $g_{H_0 qq}$ and $M_h$ with $M_{H_0}$, we have
\begin{equation}
  \Gamma^{T'}[H_0 \to gg] =
  \frac{N_c^2 \alpha_S^2 M_{H_0}^3}{576 \pi^3}
  \Bigg|
     \sum_q \frac{g_{H_0qq}}{ m_q} \left( 1 + (1-\tau_q) f(\tau_q)\right) \tau_q
  \Bigg|^2
  \,, \label{decay:Hgg}
\end{equation}
where $\tau_q=4m_q^2/M_{H_0}^2$.

\subsection{$H_0 \to \gamma\gamma$ mode}

In the standard model
the leading contribution to the $h \to \gamma \gamma$ decay width is calculated to be
\begin{equation}
 \Gamma^{\rm SM}[h \to \gamma\gamma]
 = \frac{\alpha^2 M_h^3}{256 \pi^3}
 \Bigg|
 N_c  \sum_q \frac{g_{h qq}^{\rm SM}}{\sqrt{2} m_q} Q_q^2 A_q^h(\tau_q)
 + \frac{g_{h WW}^{\rm SM}}{4 M_W^2} A_W^h(\tau_W)
 \Bigg|^2
 \,,
\end{equation}
where we neglected contributions from lepton-triangle loops, and defined~\cite{Spira:1997dg}
\begin{eqnarray}
 A_q^h(\tau) &=& 2 \tau [ 1 + (1-\tau) f(\tau) ]
\,, \\
A_W^h(\tau) &=& - [2 + 3 \tau +  3 \tau (2 - \tau) f(\tau)]
\,
\end{eqnarray}
with $\tau_i=4 m_i^2/M_h^2$.
Replacing couplings and masses with the appropriate ones, we get
\begin{equation}
 \Gamma^{T'} [H_0 \to \gamma\gamma]
 = \frac{\alpha^2 M_{H_0}^3}{256 \pi^3}
 \Bigg|
 N_c  \sum_q \frac{g_{H_0 qq}}{\sqrt{2} m_q} Q_q^2 A_q^{H_0}(\tau_q)
 + \frac{g_{H_0 WW}}{4 M_W^2} A_W^{H_0}(\tau_W)
 \Bigg|^2
 \,,
\end{equation}
  where $\tau_i=4 m_i^2/M_{H_0}^2$.

\subsection{$H_0 \to VV^*$ mode}

In the standard model
the leading contribution to the $h \to VV^*$ decay width is calculated to be
\begin{equation}
 \Gamma^{\rm SM}[h \to VV^*]
 = \delta_{V'} \frac{3 G_F (g_{h VV}^{\rm SM})^2 M_{h}}{256 \sqrt{2} \pi^3} R\left(\frac{M_V^2}{M_h^2}\right)
\,,
\end{equation}
where $G_F = \frac{1}{\sqrt{2} v_{\rm EW}^2}$ and
we defined~\cite{Spira:1997dg}
\begin{eqnarray}
\delta_{V'} &=& \Bigg\{
\begin{array}{cc}
1 & {\rm for} \, W \\
\frac{7}{12} - \frac{10}{9} s_W^2 + \frac{40}{27} s_W^4 & {\rm for} \, Z
\end{array}
\,, \\
R(x) &=&
\frac{3(1-8 x + 20 x^2)}{\sqrt{4x-1}} \cos^{-1} \left( \frac{3 x-1}{2 x^{3/2}} \right)
- \frac{(1-x)(2-13x+47x^2)}{2x}
\nonumber \\
&&
- \frac{3}{2} (1 -6x + 4x^2) \log x
\,.
\end{eqnarray}
Replacing $g_{hVV}^{\rm SM}$ and $M_h$ with $g_{H_0 VV}$ and $M_{H_0}$, respectively,
we obtain
\begin{equation}
 \Gamma^{T'}[H_0 \to VV^*]
 = \delta_{V'} \frac{3 G_F (g_{H_0 VV})^2 M_{H_0}}{256 \sqrt{2} \pi^3} R\left(\frac{M_V^2}{M_{H_0}^2}\right)
\,.
\end{equation}


\newpage

 \end{document}